\documentclass[11pt,twoside]{article}  
\usepackage{asp2006}
\usepackage{adassconf}
\begin{document}   
\paperID{8B.3}
\title{Advanced Data Reduction Techniques for MUSE}

\markboth{Weilbacher et al.}{Advanced Data Reduction Techniques for MUSE}
\author{Peter M. Weilbacher, Joris Gerssen, Martin M. Roth, Petra B\"ohm}
\affil{Astrophysikalisches Institut Potsdam, An der Sternwarte 16, D-14482 Potsdam, Germany}
\author{Arlette P\'econtal-Rousset}
\affil{Centre de Recherche Astronomique de Lyon, 9 Avenue Charles Andr\'e, 69561 Saint Genis Laval, France}
\author{and the MUSE team}

\contact{Peter Weilbacher}
\email{pweilbacher@aip.de}

\paindex{Weilbacher, P. M.}
\aindex{Gerssen, J.}
\aindex{Roth, M. M.}
\aindex{B\"ohm, P.}
\aindex{P\'econtal-Rousset, A.}


\begin{abstract}          
MUSE, a 2nd generation VLT instrument, will become the world's largest integral
field spectrograph. It will be an AO assisted instrument which, in a single
exposure, covers the wavelength range from 465 to 930\,nm with an average
resolution of 3000 over a field of view of 1'$\times$1' with 0.2'' spatial
sampling.  Both the complexity and the rate of the data are a challenge for the
data processing of this instrument.\\
We will give an overview of the data processing scheme that has been designed
for MUSE. Specifically, we will use only a single resampling step from the raw
data to the reduced data product. This allows us to improve data quality,
accurately propagate variance, and minimize spreading of artifacts and
correlated noise. This approach necessitates changes to the standard way in
which reduction steps like wavelength calibration and sky subtraction are
carried out, but can be expanded to include combination of multiple exposures.
\end{abstract}

\section{The Instrument}
MUSE is planned to be a giant integral field spectrograph and going to be
commissioned as a second generation VLT instrument starting in 2012. Its
1\arcmin$\times$1\arcmin field of view is sampled at 0\farcs2 resolution in
wide-field mode. In the adaptive-optics supported mode, observing in a
narrow-field mode with sampling at 0\farcs75 is possible.  Each exposure
delivers about 90000 spectra, using 24 slicer integral field units (IFUs), each
equipped with a separate 4k$^2$ CCD.  The full wavelength range from 465 to
930\,nm is covered with resolution between 2000 and 4000.

The light at the VLT focal plane is divided into 24 sub-fields, each of which
has a size of 60\arcsec$\times$2\farcs5 on the sky. Within each IFU, the
sub-field is then further decomposed into 48 slices. Each slice has a size of
15\arcsec$\times$$\sim$0\farcs2 on the sky. The spectrograph in each IFU
disperses the light and focuses the spectral pattern on a 4k$\times$4k CCD.
This means that the data appears on the CCDs as 48 strips of spectra which are
75\,px wide and 4096\,px high. In between these strips of data, dark regions of
$\sim$6\,px width can be used to identify stray-light.

Each raw exposure has a size of roughly 800\,MB; by using the full capabilities
of the Euro3D format, a processed exposure can require about 4.5\,GB of
storage. Typical observing nights will have 50\,GB of raw data per night
(calibrations plus science exposures), but the data rate may be as high as
150\,GB per night. Each exposure will contain approximately 350 million
illuminated pixels.

\section{The Data Processing Pipeline}
The main goals of the MUSE data processing pipeline are to fully reduce all
exposures into ready-to-use datacubes (Euro3D format or FITS with NAXIS=3)
without creating a backlog (a requirement on scope of the reduction and the
speed), to track bad pixels, propagate error information,
and to minimize the rebinning steps.
Especially the last two goals are connected and not yet commonly implemented in
existing data reduction systems. We therefore discuss the details in the
following sections.

\subsection{Noise Propagation}
We take the approach of G\"ossl \& Riffeiser (2002) and estimate the variance
from the raw data:
\begin{displaymath}
\sigma^2_\mathrm{initial} = \sigma^2_\mathrm{bias}
                          + \frac{\sigma^2_\mathrm{bias}}{n_\mathrm{bias}}
                          + \frac{\mathrm{counts} - \mathrm{bias}}{\mathrm{gain}}
\end{displaymath}
The terms describe the read-out noise, the error of the read-out noise
estimate, and the photon noise of each pixel.

This variance is then tracked through all processing steps using Gaussian error
propagation. This approach assumes that the CCD pixels are uncorrelated which
should be true to first order.  During data analysis, the variance, together
with the data value itself, makes it possible to directly read off the S/N
estimate at each position in the output datacube.

\subsection{One-step Resampling}
If we followed the classical approach to determine a calibration function for
one type of distortion and applied it (requiring resampling), before determining
the next one, MUSE data would have to be resampled four to five times.
Every resampling step always adds noise to a dataset. To be able to display and
analyze data one always needs a rectangular format, so at least one resampling
step has to be carried out.
Resampling more than once complicates the noise propagation,
passing it through five steps is near impossible.

The approach we use for the one-step resampling is centered around what we call
a ``pixel table'' (Davies 2007 describes a very similar approach that was
developed for the KMOS instrument).
This table contains a list of all illuminated pixels (the data value, the bad
pixel status, and the variance) of one exposure together with the coordinates
($x$, $y$, and $\lambda$). As soon as first estimates of the three coordinates
are known, i.\,e.\ following tracing and wavelength calibration, the CCD-based
data is transfered into such a pixel table.

Once the data are listed in the pixel table, each transformation that would
classically require an interpolation step can then be carried out just by
changing the coordinates of each table entry. Shifting the data (such as done
when combining multiple exposures) just adds a constant to the column
representing the spatial axis; tilting it (as needed to correct for
differential atmospheric refraction) applies a function to both spatial axes.
Astrometric calibration can be carried out by replacing pixel values in the two
spatial axes with values of RA and DEC (using a function describing an
astrometric solution).

To determine the function that needs to be applied to the dataset for a given
transformation one in most cases needs to resample the data already into a
datacube. An example is the flux calibration where the standard star exposure
needs to be concerted to a datacube before one can measure the flux for each
wavelength to derive the response function.
But as the intermediate datacubes are generated of auxiliary data and are not
used after the measurement, this does not affect the data quality of the
science data. Additionally, for some operations it is possible to generate the
temporary datacube with coarse resolution, and should therefore not have a
large impact on the speed of the operation when compared to the classical
approach.

%

\subsection{Final Resampling}
The final resampling involves the transformation from the pixel table to the
datacube. The problem here is that due to instrumental distortions the data are
irregularly spaced. It is difficult to find neighboring pixels needed to
interpolate.

A naive approach to this problem is to search through the table to find the
closest point(s) to be interpolated onto the output grid point. This means
to search millions of pixels repeatedly, to compute the distances to the output
grid point. For the data of only one MUSE IFU ($\sim$15 million pixels) this
process takes more than a week. This method can be sped up to only take hours
when using knowledge of the instrument geometry and a system of hierarchical
lookup tables so that one can restrict the search to a few thousands of pixels
for each output point.

Research of other existing algorithms that resample irregular data on an output
grid shows that all of them use nested loops and so work efficiently only for
sets of small size. Hence, we developed our own algorithm that only uses
``flat'' loops. In this scheme, which we call ``grid cell'' approach, we first
sort all pixels into a three-dimensional array that represents the output grid.
Each input pixel is assigned to the nearest output grid point. This means that
while in some cases none or more than one pixel are located in each grid point,
it is still easy and fast to find neighboring pixels. And if the grid is well
chosen for the input data, every grid cell will on average contain one input
pixel.

We plan to implement a few interpolation algorithms. The simplest one, nearest
neighbor, is almost trivial to implement, and the fastest possible scheme. The
whole conversion from the pixel table to the output datacube takes only about
30\,s per MUSE IFU. Another interpolation to be implemented is a distance-based
weighting method. For this, we will use a modified inverse square law for
interpolation: $f_w = \left( \frac{r_c-r}{r_c r} \right)^2$.
This function is taken from Renka (1988) and vanishes at a cutoff radius $r_c$,
so that interpolation can be restricted to the nearest 6 ($r_c=1.25$) or 26
($r_c=1.75$) grid cells.
Finally, Kriging (Krige 1951) and sinc interpolation are candidates for further
resampling techniques, but at this point it is not clear if they are usable with
MUSE data.

\section{Pipeline Implementation}
The MUSE data processing system will be implemented in the ESO context, as
plugin (``recipes'') to esorex, Gasgano, or Reflex (Hook et al., 2008). The
code will be written in C and use the Common Pipeline Library (CPL) provided
by ESO.

The pipeline will consist of a basic level of data reduction (bias, dark, and
flat-field combination, wavelength calibration, application of calibration to
science data), handling data on the basis of each integral field unit, i.e. per
CCD, separately. The second level is required to derive scientifically usable
data and will work on data from all units (across all sub-fields)
simultaneously. This part consists of a complex procedure to determine the
relative position of each slice within the field of view (the ``geometry''),
and on-sky calibration procedures like standard star reduction, handling of
separate sky and astrometric exposures, as well as combination of multiple
exposures into the final output datacube.\\
The basic level of reduction can be trivially parallelized, running multiple
reduction processes at the same time. The second level is designed to use
OpenMP to run several processing threads in parallel while processing the data
across all 24 sub-fields.

\section{Summary}
The combination of high data rate and the very complex structure makes it
difficult and possibly very time consuming to handle MUSE data.

The ``one-step resampling'' approach presented here is conceptually simple. In
this scheme it is necessary to carry out additional steps to be able to perform
the same tasks as in the classical reduction approach. CPU and memory
requirements are therefore higher, but the improved data quality and the
knowledge of variance make this a suitable method for MUSE data processing.

The final design review of the MUSE instrument is going to take place in March
2009.  Following this date, the data reduction system will be implemented to be
ready for commissioning starting end of 2012.

\acknowledgments
PMW acknowledges support by the German Verbundforschung through the MUSE/D3Dnet
project (grant 05A08BA1).
Discussions with Richard Davies and Jarle Brinchmann were important to develop
and improve the resampling algorithm.



\begin{references}
\reference Davies, R., 2007, in: The 2007 ESO Instrument Calibration Workshop, in press
\reference G\"ossl, C.~A. \& Riffeser, A., 2002, A\&A 381, 1095
\reference Hook, R., Romaniello, M., Ullgr\'en, M., et~al., 2008, The Messenger 131, 42
\reference Krige, D.~G., 1951, Master's thesis, University of Witwatersrand
\reference Renka, R., 1998 ACM 14, 151
\end{references}
\end{document}